# Featured Snippets Results in Google Web Search: An Exploratory Study

Artur Strzelecki[1] [0000-0003-3487-0971] and Paulina Rutecka[1] [0000-0002-1609-9768]

[1] University of Economics in Katowice, Department of Informatics, Katowice 40-287, Poland
{artur.strzelecki}{paulina.rutecka}@ue.katowice.pl

**Abstract.** In this paper authors analyzed 163412 keywords and results with featured snippets collected from localized Polish Google search engine. A methodology for retrieving data from Google search engine was proposed in terms of obtaining necessary data to study featured snippets. It was observed that almost half of featured snippets (48%) is taken from result on first ranking position. Furthermore, some correlations between prepositions and the most often appearing content words in keywords was discovered. Results show that featured snippets are often taken from trustworthy websites like e.g., Wikipedia and are mainly presented in form of a paragraph. Paragraph can be read by Google Assistant or Home Assistant with voice search. We conclude our findings with discussion and research limitations.

## 1  Introduction

Google is now the most important internet search engine. Google engineers are constantly developing the search algorithm to make it the most consistent with mobile standards and voice search adapted.

The mobile first idea assumes that the first step of the web design process is making a mobile view. The website must meet the requirements of the mobile user and give him or her the best experience. UX of websites is firstly designed for mobile devices and affects the desktop version (progressive enhancement). This idea arose from the fact that it is easier to add elements to desktop version than to wrap it down to the mobile version. This tendency is also correlated with the growing number of mobile users. The website has to load in a short time and does not need any further actions like taps or user decisions. In effect, websites are designed in a one-page layout that does not require any action except scrolling the screen.

The statistics show that mobile users are the largest group of web users. In 2016 it was the first time when mobile traffic was bigger than desktop traffic and amounted to 51,3% (according to statcounter.com data). According to data from Global Digital 2019 report made by We Are Social in Hootsuite in January 2019 [1] there was 5.11 billion unique mobile users, which means an increase of 100 million (2 percent) compared to

2018. In 2019 mobile devices with the result of 51.6% are on the first place among the devices through which users browse the network.

The first snippets were introduced by Google in 2012. Their goal was to show the most important information on the Search Engine Result Page (SERP). In addition to the snippets, which have been popular for several years like a knowledge graph or multimedia carousel, direct answer is becoming an increasingly common snippet.

The direct answer snippet also called the featured snippet was introduced by Google in 2016 (in Poland on April 2016). It appears as a succinct response in the form of a paragraph, list or table. The direct answer gives a highly quick response to the user query. This kind of snippet is also suitable for voice search. Voice search is becoming more popular by introducing it not only on Google Home devices but also as Google's voice assistant on mobile phones. The assistant initially debuted in May 2016 (in Poland on January 2019). Direct answer snippet is placed on the top of the search results page. It is also designed to be read by Google's Voice Assistant.

Motivation behind this study is to explore the new featured snippet that appears in web search results. Authors want to gain knowledge what factors on a keyword level and on a website level cause displaying of this snippet. The goal of our study is to analyze 163 412 keywords along with search quantity data and direct answers' source to answer the question about the correlation between types of keywords and direct answers, as well as the correlation between keywords and the answer given.

The paper is organized as follows. Section 2 contains a review of the relevant literature on snippets. Section 3 includes the methodology for data retrieval and processing, while section 4 presents the data and quantitative results. In section 5, the authors highlight the contribution of the research, discuss its limitations and, finally, draw conclusions about the results and propose possible future research avenues.

## 2 Literature review

Search engines display different snippets in SERPs. Snippets can be divided into five categories [2]: regular snippets (1), rich snippets (2), Google News (3), entity types (4) and featured snippets (5).

The first category are regular snippets displayed for typical, organic results. Since the beginning of search engines regular snippets were displayed as two lines of description placed below the title and URL of the result. [3]. For last few years it has been observed that some tests of changing length of regular snippets either on desktop or on mobile version have been done [4, 5]. Research interest for regular snippets is about their function of being enough informative for users [6]. Recent research were conducted on different age groups to determine how regular snippets are perceived by younger or older search engine users [7].

The second category are rich snippets that come from structured data dictionary schema.org [8]. Google, Microsoft (Bing), Yahoo and Yandex have founded schema.org and are recognizing structured data included in RDFa, Microdata or JSON [9]. Rich snippets created on structured data are displayed along with regular snippets [10]. Search engines show results with expanded data about products availability, price and condition, recipes, reviews, jobs, music, video and others, included in schema.org. Rich snippets are considered as important variable, especially when examining bottom-ranked results [11].

The third category are snippets displayed from Google News. These snippets are part of specialized news service and are created automatically [12]. By news publishers they are considered in different ways. Recently in Germany and Spain Google news was restricted, cause displaying snippets of news releases violates copyrights of news publishers [13]. Some propositions are made to solve this possible violation e.g., a plan for ancillary copyright is proposed, by creating original snippets [14].

The fourth category are entity types snippets. Entity types in Google are known as Knowledge Graph introduced in 2012 year and in Bing are known as Satori, launched in the same year [15]. These entities are constructed object and concepts, including people, movies, places, events, books, arts, science, etc. These entity databases are considered by users as an important part of search results [16].

The last category are featured snippets. It is the latest improvements of SERP. The search engine retrieves pieces of information from web pages and displays it in a box, above organic results together with a source URL. Google programmatically determines that a page contains a likely answer to the user's question and presents the result as a featured snippet. The other known name for this snippet is a direct answer or answer box. Direct answer supposed to deliver answers for keywords, without need to visit the result presented in search engine [17]. This snippet is displayed in four different forms like paragraph [18], table [19] and ordered or unordered list [20].

In early stage of featured snippets deployment one research examined the possible factors of presenting piece of information from particular URL in featured snippets [20]. Featured snippets was named as Google Answer Box and noticed that some prerequisites need to be met if a website could be included in the featured snippet, like: high ranking, multiple keyword inclusion in website's content, different location for keyword like headings, title, URL, paragraph, image ALT, links and structured content in the form of ordered or unordered list. Search engines creates objects displayed among search results and also they delist results because of the variety of reasons [21].

## 3  Research method

Dataset was collected using Senuto (https://www.senuto.com/). Senuto is as online tool, that retrieves data on websites visibility from Google search engine. Visibility in search engine in its basic structure is a set of three elements: keyword, ranking and result [22]. On a daily basis Senuto extracts data from Google based on the keywords list and saves results along with the position. Senuto has a database of 20 million keywords. Each keyword is at least once in a month entered to Polish localized Google search engine and a list of top 50 results is returned.

A dataset from Senuto was acquired in July 2019. Except this basic structure, authors asked Senuto owners to modify tool, to be able extracting more data from SERP. The goal of this research is to examine feature snippets, thus extended structure for extract-ing results was prepared. In extended visibility structure, along with basic set, additional elements are collected, and dataset contains the following elements: keyword, number of monthly searches, date of last extraction, parameters, featured domain, featured main domain, featured position, featured type and featured URL.

Dataset with extended visibility structured contains 163 412 examined keywords that were resulting in SERP with featured snippet. Keyword (also called as query) is the term entered to search engine interface. Keyword length is from one term up to 25 terms in case of Google. Other search engines have other maximum number of terms included in one query. Number of monthly searches is number of searches for this keyword in Google polish localized version. Data is imported through API from Google Ads. Date of last extraction is date for retrieving data.

Parameters contains internal list of snippets displayed for particular keyword. Parameters inform that for this particular keyword except feature snippet, in close proximity of feature snippet also can be displayed as: site links (1), image thumbs (2), google ads (3), map (4), brand (5), wiki on right side (6), city (7), news (8), name (9).

Site links are additional URL results displayed only for domain on first position in organic results. Image thumbs show a row with images as a piece of result from image search. Google Ads snippet present sponsored search results from Google Ads platform. Map shows a piece of map connected with the query. Brand means that for search keyword, some information can be taken from other services offered by search engine like Google My Business. Top info is displayed above featured snippet. Wiki on right contains piece of information taken from Wikipedia.

Featured domain is an exact part of URL between protocol (HTTP of HTTPS) and first slash. Featured main domain is normalized domain name. It does not contain any prefixes before the name, such a subdomain. Featured position is ranking position of results displaying the same URL that is displayed in featured snippets. Feature snippets are taken from results 1 to 10, only from SERPs first page. Featured type defines three dimensions of displayed feature snippet: paragraph, list and table. Featured URL is the final URL of the result that piece of information is displayed from.

## 4   Data and results

This section presents the data description and summary of results from data.

### 4.1   Data description

Domains, most often appearing as a source of direct answer and their position in search results, as well as links between direct answer and other snippets and the type of direct answer were analyzed.

**Table 1.** Type of featured snippet.

| featured type | frequency | percentage |
|---|---|---|
| paragraph | 114465 | 70,05% |
| list | 46509 | 28,46% |
| table | 2438 | 1,49% |

**The answer type:** Direct answers appear in three basic forms: as an article, a list and a table (Table 1.). The paragraph understood as the text response to the query appeared most often - 114298 times, which is 69.94% of the total direct answers. The second in order of frequency is the list form, which appeared in 46509 results, and that is 28.46% of the analyzed phrases. The least frequent direct answer appearing is the table which constitutes only 1.49% of results (2438 occurrences). The unclassified response type appeared 167 times, representing 0.10% of the results.

Table 2. Ranking position for featured snippet

| position | frequency | percentage |
|---|---|---|
| 0 | 485 | 0,30% |
| 1 | 79867 | 48,87% |
| 2 | 30618 | 18,74% |
| 3 | 20878 | 12,78% |
| 4 | 14469 | 8,85% |
| 5 | 9582 | 5,86% |
| 6 | 2860 | 1,75% |
| 7 | 1909 | 1,17% |
| 8 | 1319 | 0,81% |
| 9 | 860 | 0,53% |
| 10 | 554 | 0,34% |

The position of the answer's source domain: A factor, which had also been analyzed, was the position of the website which was the source from which the answer quoted as the direct answer was downloaded. The vast majority of the fragments used in direct answer came from the first 10 search results domains.

Most often are cited domains from the first search position - 79868 results were cited from these domains (48.88%). The results from the second position were cited 30618 times (18.74%), from the third position - 20878 times (12.78%). Other items are shown in the Table 2. Domains that were below the first 10 search results were cited 1 time (domain from positions 14, 15, 25, 32, 33, 41, 42) or 3 times (domain from item 11). For 485 records, a result of 0 appears, which should be classified as a data read error.

Table 3. Other snippets displayed along with featured snippet

| params | frequency | percentage |
|---|---|---|
| image thumbs | 102934 | 62,99% |
| site links | 41348 | 25,30% |
| brand | 24214 | 14,82% |
| wiki | 18675 | 11,43% |
| ads | 3148 | 1,93% |
| name | 2850 | 1,74% |
| map | 1807 | 1,11% |
| city | 1062 | 0,65% |
| news | 107 | 0,07% |

**Occurrence with other snippets:** Direct answers also occur along with other types of snippets as shown in Table 3. They appear most often in the company of an image -

102934 times (62.99%) or site links - 41348 times (25.30%). Less frequently, along with featured snippets, brand - 24214 times (14.82%), wiki (knowledge graph) - 18675 times (11.43%). The other snippets shown in the Table 3. that appear together with the direct answer are marginal. These are Google Ads (1.93%), name (1.74%), map (1.11%), city (0.65%) and news (0.07%). A direct answer alone with no other snippets appears 26813 times, which is 16.41% of the analyzed results. Brands are keywords, which contain a term with a brand name and are recognize based on site links, e.g., brand keyword is "refrigerator samsung".

**Source domains:** Among the emerging domains cited by Google in direct answer the most-appearing is wikipedia.org - internet encyclopedia (48242 times), medonet.pl - health category (2352 times), fandom.com - entertainment (1259 times), kwestiasmaku.com - cooking (1033 times), apteline.pl - online pharmacy (941 times), sciaga.pl - education (913 times), goodreads.com - books (847 times), quora.com - Q&A service (835 times), chillizet .pl - radio information service (833 times) and mp.pl - health (762 times).

### 4.2   Results

The analysis of the possessed data allowed to divide the keywords into 3 main sections due to the syntax of the key phrase. First one contains keywords started with pronouns or containing pronouns characteristic for the syntax of the interrogative sentence. The second section contains keywords beginning with or containing an attribute such as: cost, weight, capacity. The third section of keywords contains phrases beginning with or containing an adverb's or adjective's superlative form.

**Table 4.** TOP pronouns in keywords

| pronouns | frequency of keywords beginning with | frequency of keywords containing |
| --- | --- | --- |
| co | 3029 | 4604 |
| jak | 2724 | 1328 |
| ile | 843 | 670 |
| kiedy | 674 | 365 |
| czy | 106 | 295 |
| kim | 182 | 229 |
| dlaczego | 488 | 220 |
| kim | 379 | 131 |
| dlaczego | 187 | 123 |
| jakie | 220 | 90 |
| jaka | 148 | 87 |
| kim | 64 | 30 |
| czy | 3 | 16 |
| dlaczego | 79 | 14 |

**Keywords syntax – pronouns:** From the keywords there were separated pronouns as follows: what, how, when, how much/how many, who, where, why, with what, whom, which. Among the analyzed 163 412 keywords, 9131 starts with one of the

words mentioned above (Table 4.), which is 5.59% of the keywords analyzed in general. 8192 analyzed key phrases contain the pronoun, but it does not appear as the first word in the phrase. This is 5.01% of the entire analyzed group of keywords.

Attributes of keywords content: Queries often refer to attributes that define a feature. From the analyzed keywords, these keywords were distinguished, which contain the attribute. 89 phrases which are the query about the object's, product's, country's or person's feature has been separated. The Table 5 shows the 10 most often used keywords and 10 attributes that appear in a keyword, but not on the first position.

Table 5. TOP 10 attributes

| keywords beginning with attribute | frequency | keywords contain attribute | frequency |
|---|---|---|---|
| objawy | 1143 | objawy | 3164 |
| przepis | 665 | definicja | 2550 |
| rodzaje | 664 | cena | 1403 |
| definicja | 356 | przepis | 1140 |
| wymiary | 307 | wymagania | 906 |
| cena | 273 | znaczenie | 841 |
| wymagania | 271 | wymiary | 547 |
| temperatura | 202 | rodzaje | 420 |
| znaczenie | 160 | kalorie | 411 |

89 separated attribute definitions appear as the first word in key phrase for 10173 phrases, which accounts for 6.23% of the analyzed keywords. In the next positions there are up to 18050 key phrases, which accounts for 11.05% of the key words analyzed.

Table 6. TOP 10 superlatives

| keywords beginning with superlative | frequency | keywords contain superlative | frequency |
|---|---|---|---|
| najładniejsze | 210 | najpiękniejsze | 43 |
| najmłodsza | 190 | najwięksi | 38 |
| najdroższe | 121 | najładniejsze | 20 |
| najpiękniejsze | 117 | najdroższa | 18 |
| najwięksi | 104 | najlepsza | 18 |
| najlepsza | 69 | najwięcej | 16 |
| najdroższa | 60 | najmłodsza | 14 |
| najdłuższy | 60 | najważniejsze | 14 |
| najdłuższa | 59 | najlepiej | 14 |

Keywords with superlative prefix (Table 6.): In the analyzed keywords, 202 keywords with the superlative prefix "naj" (suggesting the best or the worst) were extracted. 2401 key phrases started with the words 'the best or the worst', which is 1.47% of the words analyzed in general. 378 contained these phrases on a different then the first position of the phrase (0.23% of the analyzed keywords). The same words appear in the most popular list, regardless of whether it is at the beginning or elsewhere in the key phrase. Definitely more often the words the best and the worst appear at the beginning of the key phrase formation. The analysis at the level of individual URLs indicates

that the majority of the phrases that result in the display of featured snippet refers to queries for culinary, medicine and IT.

## 5    Conclusion and Discussion

In this paper, we presented an analysis of the data set that causes appearance of featured snippets (direct answer or Google answer) in the search engine. The findings of our study indicate that the Google search engine is being developed in the direction of displaying the query response from the search results page. Google does not discriminate blue links, but makes the valuable site stand out. Because there is no automatic inclusion and no mark that can be added to website for being taken into consideration for featured snippets displaying, our findings on a keyword level help to prepare better content on the website, to be considered for this snippet.

The analysis of data shows that the appearance of a direct answer is closely related to the question form of the phrase (the occurrence of pronouns) or the occurrence of words in the phrase specifying an attribute that has a specific value, considered as the answer to the query, e.g. product price. Frequently appearing phrases are also queries containing adjectives in superlative form, e.g. the highest peak, the largest city. All these queries have unambiguous and undeniable value.

Google uses as a source from which direct answer is cited only pages with high ranking in search engines which meet the criteria of Google's ranking factors. Meeting these criteria defines the site as containing content of a high value, popular among users. This means that the website enjoys the trust of both the search engine and the users themselves. This makes it very likely that the information in the direct answer is correct. The same websites, such as Wikipedia, are often cited.

The most common types of direct answers are responses in the form of a paragraph. This kind of answer is the most legible and at the same time the most convenient to be read by voice search systems. It potentially allows the computer for a voice response using a speech synthesizer. Results in the form of a list or table appear less frequently. The form of a list usually appears in cookery recipes (ingredients for preparing a dish) or in medicine, when symptoms of a disease are mentioned. The table appears most frequently for inquiries regarding prices or other values related to financial products (such as: taxes, loans, insurance). Featured snippets often occur along with other types of snippets, of which the most common are images that draw the user's attention by highlighting themselves in the search results list, as well as links to the pages from which the data used for the answer come. By correlation with links to direct pages, they're called zero position in search results, and SEO specialists are analyzing how one can meet enough ranking factors to make the page cite the direct answer. This item is very often clicked, and the page in this position gains the position of an expert.

The limitation of our research was the fact of having a set of data concerning 163 412 keywords only in Polish. All data concern the Google search engine, which is dominant in Poland. However, we realize that results containing featured snippets types are observed in other language versions of Google as well. Other limitation is data dataset is relatively small when comparing it to volume of searches made daily by search engine users.

Future research will be conducted to investigate the factors affecting the display of results from specific websites in the snippets area. Another direction of future studies is to analyze content of featured snippet by extracting paragraphs, tables and lists from SERPs and study them. However, this direction demands to extend even more, already extended data set for studying visibility in search engines.